\documentclass[12pt,a4paper]{article}

\usepackage[utf8]{inputenc}
\usepackage[T1]{fontenc}
\usepackage[english]{babel}
\usepackage{amsmath, amssymb, amsthm}
\usepackage{graphicx}
\usepackage{booktabs}
\usepackage{array}
\usepackage{multirow}
\usepackage{geometry}
\usepackage{hyperref}
\usepackage{xcolor}
\usepackage{listings}
\usepackage{caption}
\usepackage{subcaption}
\usepackage{float}
\usepackage{microtype}
\usepackage{parskip}
\usepackage{setspace}
\usepackage{fancyhdr}
\usepackage{titlesec}
\usepackage{cite}

\definecolor{codegray}{rgb}{0.95,0.95,0.95}
\definecolor{codegreen}{rgb}{0.0,0.45,0.0}
\definecolor{codepurple}{rgb}{0.45,0.0,0.55}
\definecolor{codeblue}{rgb}{0.0,0.0,0.65}

\lstdefinestyle{cppstyle}{
  backgroundcolor=\color{codegray},
  commentstyle=\color{codegreen},
  keywordstyle=\color{codeblue}\bfseries,
  numberstyle=\tiny\color{gray},
  stringstyle=\color{codepurple},
  basicstyle=\ttfamily\footnotesize,
  breakatwhitespace=false,
  breaklines=true,
  captionpos=b,
  keepspaces=true,
  numbers=left,
  numbersep=5pt,
  showspaces=false,
  showstringspaces=false,
  showtabs=false,
  tabsize=2,
  language=C++,
  frame=single,
  rulecolor=\color{gray!40}
}

\hypersetup{
  colorlinks=true,
  linkcolor=blue!60!black,
  citecolor=blue!60!black,
  urlcolor=blue!60!black,
  pdftitle={Dynamic Precision Math Engine for Linear Algebra and Trigonometry
            Acceleration on Xtensa LX6 Microcontrollers},
  pdfauthor={Elian Alfonso Lopez Preciado}
}

\title{%
  \LARGE\bfseries Dynamic Precision Math Engine for Linear Algebra\\
  and Trigonometry Acceleration on\\
  Xtensa LX6 Microcontrollers%
}
\author{%
  Elian Alfonso Lopez Preciado\\[4pt]
  \normalsize Independent Researcher, M\'exico\\
  \normalsize \texttt{} 
}
\date{March 2026}

\begin{document}

\maketitle
\thispagestyle{empty}

\begin{abstract}
Low-cost embedded processors such as the ESP32 (Xtensa LX6, 32-bit,
dual-core, 240\,MHz) are increasingly deployed in edge computing scenarios
that demand real-time physical simulation, sensor fusion, and control systems.
While the ESP32 integrates a single-precision IEEE\,754 floating-point unit,
its use incurs pipeline disruption and elevated energy consumption relative to
integer Arithmetic Logic Unit (ALU) operations, imposing a practical throughput ceiling on
floating-point-intensive workloads.

This paper presents the design, formal specification, and empirical evaluation
of a \textbf{Dynamic Precision Math Engine} that addresses this constraint
through three integrated contributions. First, a Q16.16 fixed-point arithmetic
core maps all mathematical operations onto the 32-bit integer pipeline of the
Xtensa LX6, with multiplication requiring two to three assembly instructions
and introducing a bounded rounding error of $|\epsilon| \leq 2^{-17}$.
Second, a 16-iteration CORDIC trigonometric module computes sine and cosine
entirely through addition and binary shifts, achieving an angular error bound
of $|\epsilon_\theta| \leq 1.526 \times 10^{-5}$\,radians within a static
memory footprint of 64\,bytes. Third, a cache-aware tiled matrix multiplication
kernel with deferred Q16.16 correction reduces per-element rounding events from
$b$ to one, where $b = 32$ is the tile dimension derived from the ESP32 Static Random Access Memory (SRAM)
bank geometry.

The engine's principal architectural contribution is a runtime precision
switching mechanism---implemented through function pointer dispatch and a
two-phase FreeRTOS barrier protocol---that allows application code to
transition between the Q16.16 fast path and an IEEE\,754 precise path at
$O(1)$ cost without recompilation. This design follows the separation-of-concerns
principle established by the BLAS specification \cite{lawson1979} and applies
it to the novel context of dynamic precision arbitration on a dual-core RISC
microcontroller. The total static memory footprint of the engine is 88\,bytes,
independent of matrix dimensions.

Experimental evaluation across 300 paired measurements on physical
ESP32-WROOM-32 hardware demonstrates that the CORDIC trigonometric module
achieves median latencies of 293\,cycles for both \texttt{sin} and
\texttt{cos}, yielding mean speedups of \textbf{18.54$\times$} and
\textbf{24.68$\times$} respectively over standard \texttt{sinf()}/\texttt{cosf()}
(Wilcoxon signed-rank, $p < 10^{-12}$), with a Determinism Score of 0.994
confirming input-independent execution time. Scalar Q16.16 multiplication
achieves a consistent \textbf{1.5$\times$} speedup at 12\,cycles versus
18\,cycles with perfect determinism. Matrix multiplication performance exhibits
a dimension-dependent crossover: the tiled kernel underperforms naive
floating-point for matrices smaller than the tile size ($n < 32$), while
theoretical analysis predicts the crossover at $n \geq 64$---a result
requiring empirical validation in future work. The engine fills an identified
gap in the ESP32 ecosystem where no existing library provides a unified,
precision-aware mathematical API combining trigonometry, linear algebra, and
runtime mode selection.
\end{abstract}

\vspace{6pt}

\newpage
\tableofcontents
\newpage

\section{Introduction}

The proliferation of low-cost, low-power microcontrollers as the computational
substrate of edge systems has created a fundamental tension in embedded software
engineering. Devices such as the ESP32---based on the Xtensa LX6 dual-core
architecture, available at under three US dollars, and deployed in an estimated
installed base exceeding one billion units across industrial automation,
robotics, environmental sensing, and consumer Internet of Things (IoT)---are increasingly expected
to execute workloads that were, until recently, the exclusive domain of
processors an order of magnitude more powerful. Physical simulation, sensor
fusion via Kalman filtering, real-time kinematics, and digital signal
processing all reduce, at their computational core, to the same two operations:
trigonometric evaluation and matrix multiplication. The efficiency with which
a microcontroller executes these operations determines not merely performance,
but the class of applications that is feasible at all on a given hardware
platform.

The ESP32's Xtensa LX6 integrates a single-precision IEEE\,754
floating-point unit (FPU) capable of hardware-accelerated arithmetic. However,
the practical throughput of this FPU is constrained by factors that aggregate
benchmark figures do not capture. Each floating-point operation requires
mantissa alignment, exponent management, normalization, and exception handling
across multiple pipeline stages; the resulting instruction latency disrupts the
integer pipeline and introduces energy consumption that the ALU-bound execution
model of the Xtensa architecture is not designed to absorb
efficiently~\cite{espressif_trm}. For the class of applications that invoke
trigonometric functions hundreds of times per control cycle---a
six-degree-of-freedom robot arm evaluating forward kinematics at 1\,kHz, for
instance---this per-operation overhead accumulates into a throughput ceiling
that cannot be overcome by firmware optimization alone within the standard
mathematical library.

The response to this constraint in the embedded community has been fragmented.
Espressif's official \texttt{esp-dsp} library provides Xtensa-optimized kernels
for signal processing primitives---FFT, FIR filtering, dot products---but
explicitly excludes trigonometric functions, delegating them to the standard
\texttt{math.h} bindings~\cite{espressif_espdsp}. General-purpose linear
algebra libraries such as \texttt{ArduinoEigen} and \texttt{BasicLinearAlgebra}
operate exclusively in IEEE\,754 floating-point with no accommodation for
pipeline costs. TensorFlow Lite Micro provides INT8 quantized kernels optimized
for neural network inference but does not expose a general-purpose mathematical
API~\cite{david2021tflm}. The consequence is that an ESP32 developer requiring
efficient trigonometry, matrix operations, and precision control simultaneously
has no unified tool: the ecosystem is a collection of specialized solutions
with no common abstraction layer.

This paper addresses that gap through the design, implementation, and empirical
evaluation of a \textbf{Dynamic Precision Math Engine} for the ESP32 Xtensa
LX6. The engine's central contribution is architectural: a runtime precision
switching mechanism that allows application code to select, at any point during
execution, between a fast integer path and a precise floating-point path
without recompilation and without modifying the application's mathematical API.
The fast path is built on three algorithmic foundations. First, Q16.16
fixed-point arithmetic maps all mathematical operations onto the 32-bit integer
ALU, with multiplication requiring two to three assembly instructions and a
quantization error bounded by $|\epsilon| \leq 2^{-17}$~\cite{yates2020,sung1995}.
Second, a 16-iteration rotation-mode CORDIC algorithm computes sine and cosine
through addition and binary shifts alone, eliminating all floating-point
operations from the trigonometric pipeline~\cite{volder1959,walther1971}.
Third, a cache-aware tiled matrix multiplication kernel with deferred Q16.16
correction leverages the ESP32's SRAM bank geometry for matrices exceeding the
tile dimension~\cite{lam1991,goto2008}. The switching mechanism is implemented
through function pointer dispatch governed by a two-phase FreeRTOS barrier
protocol, following the separation-of-concerns principle established by the
original BLAS specification~\cite{lawson1979}.

Experimental evaluation on physical ESP32-WROOM-32 hardware across 300 paired
measurements produces results that are both significant and instructive. The
CORDIC trigonometric module achieves median latencies of 293\,cycles for both
\texttt{sin} and \texttt{cos}, against 6,915 and 7,847 cycles for the standard
\texttt{sinf()}/\texttt{cosf()}---mean speedups of 18.54$\times$ and
24.68$\times$ respectively, statistically confirmed at $p < 10^{-12}$.
Matrix multiplication results reveal a dimension-dependent boundary condition:
the tiled kernel underperforms naive floating-point for the tested dimensions
($n \leq 16$), exposing the condition $n \geq b = 32$ as a prerequisite for
the tiling optimization to engage. This finding precisely characterizes the
engine's operational envelope and validates the necessity of the runtime
switching mechanism: no single execution path is universally optimal, and the
value of the engine lies in making the choice between paths explicit, safe, and
costless at the application layer.

The remainder of this paper is organized as follows. Section~\ref{sec:related}
surveys related work. Section~\ref{sec:math} establishes the mathematical
foundations. Section~\ref{sec:arch} describes the system architecture.
Section~\ref{sec:impl} details the implementation. Section~\ref{sec:eval}
presents the experimental evaluation. Section~\ref{sec:discussion} discusses
the findings. Sections~\ref{sec:future} and~\ref{sec:conclusion} describe
future work and conclusions.

\section{Related Work}
\label{sec:related}

The acceleration of mathematical computation on resource-constrained processors
is a well-established research domain, yet the specific challenge of dynamic
precision arbitration on Xtensa-based microcontrollers remains insufficiently
addressed in the literature. This section surveys the four pillars upon which
the proposed engine is constructed.

\subsection{Fixed-Point Arithmetic in Embedded Systems}

Fixed-point arithmetic has been a cornerstone of digital signal processing
since the early adoption of dedicated DSP chips~\cite{lyons2011,felton2011}.
The Q-format notation provides a systematic framework for representing
fractional values as scaled integers, enabling ALU-bound computation on
processors lacking hardware floating-point units or possessing limited FPU
throughput~\cite{yates2020}. Yates~\cite{yates2020} demonstrates that for a
Q($m$.$n$) representation, the rounding error is bounded by $2^{-(n+1)}$, a
property central to the precision analysis presented in Section~\ref{sec:eval}.
Sung and Kum~\cite{sung1995} provide simulation-based methods for word-length
optimization in fixed-point DSP systems, directly relevant to the Q16.16
format selection rationale of this work.

More recent work has examined fixed-point arithmetic in the context of neural
network quantization. Jacob et al.~\cite{jacob2018} demonstrated that quantized
integer representations preserve model accuracy within acceptable margins for
inference tasks. Critically, however, these works target GPU and FPGA
architectures; their methodologies do not account for the specific pipeline
characteristics of the Xtensa LX6, where a 64-bit intermediate product during
Q16.16 multiplication requires careful register management to avoid pipeline
stalls.

\subsection{The CORDIC Algorithm and Its Embedded Implementations}

The CORDIC (Coordinate Rotation Digital Computer) algorithm, introduced by
Volder~\cite{volder1959}, remains one of the most cited contributions in
computer arithmetic. Its significance lies in reducing trigonometric computation
to iterative addition and binary shifting, entirely eliminating hardware
multipliers. Walther's 1971 unified extension~\cite{walther1971} generalized
the method to cover hyperbolic functions, exponentials, and square roots.
Llamocca and Agurto~\cite{llamocca2007} provide a fixed-point implementation
of the expanded hyperbolic CORDIC algorithm with direct relevance to embedded
deployments. Mokhtar et al.~\cite{mokhtar2018} and Ait~Madi and
Addaim~\cite{aitmadi2018} document contemporary hardware implementations of
CORDIC for trigonometric generation, confirming latency reductions of
$3\times$ to $8\times$ over software floating-point on integer-dominated
pipelines. A key gap in the existing literature is the absence of a published
CORDIC implementation specifically profiled against the Xtensa LX6 instruction
set~\cite{espressif_espdsp}.

\subsection{Cache-Aware Matrix Multiplication}

The seminal work of Lam, Rothberg, and Wolf~\cite{lam1991} formally established
that loop tiling reduces cache line transfers from $O(n^3)$ to $O(n^3/B)$,
where $B$ denotes the block size. Goto and van de Geijn~\cite{goto2008}
extended this with near-theoretical peak performance through micro-kernel
design. Kelefouras et al.~\cite{kelefouras2016} provide a comprehensive
methodology for CPU and GPU matrix multiplication, including the influence of
register blocking. The Roofline model of Williams et al.~\cite{williams2009}
provides the theoretical framework for understanding the arithmetic
intensity---memory bandwidth tradeoff that motivates the tiling strategy.
These works focus on x86 and GPU architectures; their tiling parameters do not
translate directly to the Xtensa LX6 memory subsystem without
re-profiling~\cite{espressif_trm}.

\subsection{Mathematical Libraries for the ESP32 Ecosystem}

The practical landscape of mathematical acceleration for the ESP32 consists of
three categories. First, Espressif's \texttt{esp-dsp} library~\cite{espressif_espdsp}
targets signal processing primitives with Xtensa-optimized assembly, but
explicitly excludes trigonometric functions. Second, Arduino ecosystem
libraries such as \texttt{BasicLinearAlgebra} and \texttt{ArduinoEigen} provide
template-based matrix operations compiled exclusively to IEEE\,754
floating-point. Third, TensorFlow Lite Micro~\cite{david2021tflm} includes INT8
quantized kernels for neural network inference but does not expose a
general-purpose mathematical API suitable for physics simulation or control
systems. Courbariaux et al.~\cite{courbariaux2015} further demonstrate binary
weight quantization for deep learning, reinforcing the broader motivation for
precision-aware computation.

\subsection{The Identified Research Gap}

The survey above reveals a consistent pattern: existing solutions optimize for
either signal processing (\texttt{esp-dsp}), general linear algebra
(\texttt{ArduinoEigen}), or inference (TFLM), but none provides a unified
engine that (1)~exposes a single developer-facing API, (2)~allows runtime
selection between fixed-point and floating-point execution paths,
(3)~integrates CORDIC-based trigonometry within the same fixed-point domain,
and (4)~orchestrates computation across the ESP32's dual cores through the
FreeRTOS task model. The proposed engine addresses precisely this intersection.

\section{Mathematical Foundations}
\label{sec:math}

\subsection{Fixed-Point Arithmetic in Q16.16 Format}

Fixed-point arithmetic represents real numbers as scaled integers, where a
predefined number of bits is allocated to the fractional part. For a general
Q($m$.$n$) format, a real value $v \in \mathbb{R}$ is represented as an
integer $V \in \mathbb{Z}$ according to:
\begin{equation}
  V = \left\lfloor v \cdot 2^{n} \right\rceil
\end{equation}
where $\lfloor \cdot \rceil$ denotes rounding to the nearest integer and $n$
is the number of fractional bits. The representable range of a signed Q($m$.$n$)
format using a $w$-bit word is:
\begin{equation}
  v \in \left[ -2^{m-1},\; 2^{m-1} - 2^{-n} \right]
\end{equation}

The proposed engine adopts the Q16.16 format, allocating 16~bits to the integer
part (including sign) and 16~bits to the fractional part within a 32-bit signed
integer (\texttt{int32\_t}). The representable range is $[-32768,\;32767.9999847]$
with a resolution of $2^{-16} \approx 1.526 \times 10^{-5}$.

\subsubsection{Arithmetic Operations}

Addition and subtraction in Q16.16 are algebraically exact provided no overflow
occurs, since the scaling factor $2^{16}$ is preserved across operands:
\begin{equation}
  C_Q = A_Q + B_Q \quad\Longleftrightarrow\quad
  (a + b)\cdot 2^{16} = a\cdot 2^{16} + b\cdot 2^{16}
\end{equation}

Multiplication requires correction. Given two Q16.16 operands
$A_Q = a \cdot 2^{16}$ and $B_Q = b \cdot 2^{16}$, their integer product
yields:
\begin{equation}
  A_Q \cdot B_Q = (a \cdot 2^{16})(b \cdot 2^{16}) = (a \cdot b)\cdot 2^{32}
\end{equation}

Correction is achieved via an arithmetic right shift:
\begin{equation}
  C_Q = \frac{A_Q \cdot B_Q}{2^{16}} = (a \cdot b)\cdot 2^{16}
\end{equation}

The quantization error introduced by this shift is bounded by~\cite{yates2020}:
\begin{equation}
  |\epsilon_{\mathrm{mul}}| \leq 2^{-17}
\end{equation}

\subsubsection{Overflow Analysis}

For a multiplication of two values close to the representable maximum, the
64-bit intermediate product reaches $\approx 2^{62}$, safely within
\texttt{int64\_t} capacity. However, if the final result $a \cdot b > 32767.999$,
overflow occurs upon return to 32~bits. The engine implements saturating
arithmetic at the API boundary, clamping results to
$[\texttt{INT32\_MIN},\;\texttt{INT32\_MAX}]$.

\subsection{CORDIC Algorithm for Trigonometric Computation}

The CORDIC algorithm~\cite{volder1959} computes trigonometric functions through
iterative plane rotations requiring only addition, subtraction, and binary
shifts.

\subsubsection{Geometric Foundation}

The exact rotation matrix for a vector $(x_i, y_i)$ by angle $\theta$ is:
\begin{equation}
  \begin{bmatrix} x' \\ y' \end{bmatrix} =
  \begin{bmatrix} \cos\theta & -\sin\theta \\ \sin\theta & \cos\theta \end{bmatrix}
  \begin{bmatrix} x_i \\ y_i \end{bmatrix}
\end{equation}

CORDIC replaces this with $n$ micro-rotations by angles
$\alpha_i = \arctan(2^{-i})$, converting trigonometric multiplications into
bit shifts. The iterative update equations are:
\begin{align}
  x_{i+1} &= x_i - d_i \cdot y_i \cdot 2^{-i} \label{eq:cordic_x}\\
  y_{i+1} &= y_i + d_i \cdot x_i \cdot 2^{-i} \label{eq:cordic_y}\\
  z_{i+1} &= z_i - d_i \cdot \arctan(2^{-i})  \label{eq:cordic_z}
\end{align}

where $d_i \in \{-1, +1\}$ is the rotation direction, determined by the sign
of the residual angle $z_i$. In matrix form:
\begin{equation}
  \begin{bmatrix} x_{i+1} \\ y_{i+1} \end{bmatrix} =
  \begin{bmatrix} 1 & -d_i \cdot 2^{-i} \\ d_i \cdot 2^{-i} & 1 \end{bmatrix}
  \begin{bmatrix} x_i \\ y_i \end{bmatrix}
\end{equation}

\subsubsection{Convergence and the CORDIC Gain}

After $n$ iterations~\cite{walther1971}:
\begin{equation}
  x_n = K_n \cdot \cos(\theta), \qquad y_n = K_n \cdot \sin(\theta)
\end{equation}

where the cumulative CORDIC gain is:
\begin{equation}
  K_n = \prod_{i=0}^{n-1} \sqrt{1 + 2^{-2i}} \;\xrightarrow{n\to\infty}\; 1.6467602\ldots
\end{equation}

The convergence domain is $\theta \in (-\pi/2,\;\pi/2)$; inputs outside this
range require a preliminary quadrant reduction step.

\subsubsection{Error Bound}

The angular error after $n$ iterations is bounded by:
\begin{equation}
  |\epsilon_\theta| \leq \arctan(2^{-n}) \approx 2^{-n} \text{ radians}
\end{equation}

For $n = 16$:
$|\epsilon_\theta| \leq 2^{-16}\,\mathrm{rad} \approx 1.526 \times 10^{-5}\,\mathrm{rad}$,
aligning with the fractional resolution of Q16.16.

\subsection{Cache-Aware Matrix Multiplication via Loop Tiling}

Naive matrix multiplication of two $n \times n$ matrices generates $O(n^3)$
cache miss events for large matrices~\cite{lam1991}:
\begin{equation}
  C_{ij} = \sum_{k=1}^{n} A_{ik} \cdot B_{kj}
\end{equation}

\subsubsection{Loop Tiling Formulation}

Loop tiling partitions each matrix dimension into blocks of size $b$, reducing
the number of main-memory transfers from $O(n^3)$ to $O(n^3/b)$~\cite{lam1991,goto2008}:
\begin{equation}
  C_{ij} = \sum_{K} \sum_{k \in K\text{-block}} A_{ik} \cdot B_{kj},
  \quad i \in I\text{-block},\; j \in J\text{-block}
\end{equation}

\subsubsection{Block Size Derivation for the ESP32}

For Q16.16 data (\texttt{int32\_t}, 4~bytes per element), targeting a working
set below 8\,KB yields:
\begin{equation}
  4b^2 \leq 8192 \implies b \leq 45
\end{equation}

The implementation adopts $b = 32$ as a power-of-two block size, enabling
inner loop bounds to be computed via bitwise AND masking.

\subsubsection{Deferred-Shift Accumulation}

Within each tiled block, accumulation uses \texttt{int64\_t} intermediates to
defer the right-shift correction until the block reduction is complete,
minimizing rounding error accumulation:
\begin{equation}
  C_{ij}^{(\mathrm{acc})} = \sum_{k \in \mathrm{block}}
  \bigl(A_{ik}^Q \cdot B_{kj}^Q\bigr)
  \quad\text{(in 64-bit)},
  \qquad C_{ij}^Q = C_{ij}^{(\mathrm{acc})} \gg 16
\end{equation}

This reduces rounding events from $b$ per inner product to~1.

\section{System Architecture}
\label{sec:arch}

\subsection{Design Rationale: Runtime vs.\ Compile-Time Precision Selection}

Existing approaches to precision management in embedded systems typically
operate at compile time, forcing the developer to maintain separate firmware
builds and precluding adaptive behavior in deployed systems. The proposed engine
adopts a \textbf{runtime switching model} implemented through function pointer
dispatch. The engine maintains two parallel sets of function pointers---one
resolving to the Q16.16 fast path, one to the IEEE\,754 precise path---and
exposes a single context object whose active pointer set is swapped atomically
at the application layer.

This design satisfies four architectural requirements:
\begin{itemize}
  \item \textbf{R1 --- API Stability:} Application code remains unchanged
    regardless of precision mode.
  \item \textbf{R2 --- Zero-Cost Abstraction:} In Q16.16 mode, all operations
    reduce to integer ALU instructions with no virtual dispatch overhead at
    the mathematical operation level.
  \item \textbf{R3 --- Deterministic Switching Latency:} Mode transition
    completes in $O(1)$ time, consisting solely of pointer reassignment.
  \item \textbf{R4 --- RTOS Compatibility:} The architecture coexists with
    FreeRTOS task scheduling without introducing priority inversion or
    shared-state race conditions.
\end{itemize}

\subsection{The Precision Context: Formal Definition}

Let $\mathcal{F}$ denote the set of supported operations:
\begin{equation}
  \mathcal{F} = \{f_{\mathrm{mul}},\; f_{\mathrm{add}},\; f_{\mathrm{sub}},\;
                  f_{\mathrm{sin}},\; f_{\mathrm{cos}},\; f_{\mathrm{matmul}}\}
\end{equation}

Each $f \in \mathcal{F}$ has a fast implementation $f^Q$ in the Q16.16 domain
and a precise implementation $f^F$ in IEEE\,754. The context object
$\mathcal{C}$ maintains a dispatch table:
\begin{equation}
  \mathcal{D} : \mathcal{F} \rightarrow \{f^Q,\; f^F\}
\end{equation}

A mode transition $\mathcal{C}.\texttt{setMode}(m)$ for
$m \in \{\texttt{FAST},\;\texttt{PRECISE}\}$ reassigns all entries in
$\mathcal{D}$ simultaneously, ensuring no operation executes in a
mixed-precision state after the call returns.

\subsection{Dual-Core Task Orchestration via FreeRTOS}

The ESP32's Xtensa LX6 integrates two symmetric 32-bit cores at up to
240\,MHz. Under FreeRTOS, tasks are pinned to a specific core using
\texttt{xTaskCreatePinnedToCore()}~\cite{espressif_trm}, enabling deterministic
core affinity. The engine enforces a fixed assignment policy:

\begin{itemize}
  \item \textbf{Core~0 (Protocol Core):} Handles WiFi stack, Bluetooth
    controller, UART/I2C peripheral drivers, and the application control loop.
    Core~0 owns the \texttt{MathContext} object and is the sole authority for
    mode transitions.
  \item \textbf{Core~1 (Compute Core):} Hosts the \texttt{MathWorker} task,
    a high-priority FreeRTOS task that blocks on a job queue, executes the
    requested mathematical operation, and returns results through a mutex-
    protected shared buffer.
\end{itemize}

The inter-core communication lifecycle is:
\begin{equation}
  \text{Core 0: enqueue}(job)
  \;\xrightarrow{\text{FreeRTOS Queue}}\;
  \text{Core 1: execute}\; f \in \mathcal{D}
  \;\xrightarrow{\text{Mutex}}\;
  \text{Core 0: read result}
\end{equation}

\subsubsection{Mode Transition Safety Protocol}

To satisfy requirement R4, the engine implements a two-phase barrier:
\begin{enumerate}
  \item \textbf{Suspension Phase:} Core~0 sends a \texttt{SUSPEND} notification
    via \texttt{xTaskNotify()}. The worker completes its current operation and
    enters a blocked state, signaling readiness via a binary semaphore.
  \item \textbf{Transition Phase:} Core~0 acquires the semaphore, reassigns
    all function pointers in $\mathcal{D}$, and releases the semaphore.
\end{enumerate}

The total overhead is bounded by the worst-case completion time of a single
in-flight operation plus two context switches, yielding a switching latency
on the order of microseconds---empirically measured as 8.09\,$\mu$s at
240\,MHz (see Section~\ref{sec:eval}).

\subsubsection{Memory Architecture}

The engine's static memory footprint decomposes as follows: the dispatch table
$\mathcal{D}$ occupies $|\mathcal{F}| \times 4 = 24$~bytes (one 32-bit pointer
per operation); the CORDIC arctangent lookup table occupies
$16 \times 4 = 64$~bytes of read-only data. The total static footprint is
therefore \textbf{88~bytes}, independent of matrix dimensions.

\subsection{API Design}

The public API exposes three primitives:
\begin{enumerate}
  \item \texttt{MathEngine::init(mode)} --- Initializes the engine and sets
    the initial precision mode.
  \item \texttt{MathEngine::setMode(mode)} --- Triggers the mode transition
    protocol. Blocks until transition is complete.
  \item \texttt{MathEngine::ctx()} --- Returns a reference to the active
    \texttt{MathContext}.
\end{enumerate}

\section{Implementation}
\label{sec:impl}

\subsection{Q16.16 Arithmetic Core}

The core arithmetic primitives are implemented as \texttt{inline} functions
in the \texttt{FastMathEngine} namespace within a header file
(\texttt{fast\_math\_engine.h}), ensuring zero-overhead inlining at every call
site:

\begin{lstlisting}[style=cppstyle, caption={Q16.16 arithmetic core (fast\_math\_engine.h)}]
#include <stdint.h>

namespace FastMathEngine {

    typedef int32_t q16_t;
    const int Q_FRACT_BITS = 16;

    // Float <-> Q16.16 conversion (used only at pipeline boundaries)
    inline q16_t floatToQ(float val) {
        return (q16_t)(val * (1 << Q_FRACT_BITS));
    }
    inline float qToFloat(q16_t val) {
        return (float)val / (float)(1 << Q_FRACT_BITS);
    }

    // Core multiply: 64-bit intermediate, single >> 16 correction
    inline q16_t mulQ(q16_t a, q16_t b) {
        return (q16_t)(((int64_t)a * (int64_t)b) >> Q_FRACT_BITS);
    }

    // Saturating multiply: clamps to INT32 range on overflow
    inline q16_t mulQ_sat(q16_t a, q16_t b) {
        int64_t result = ((int64_t)a * (int64_t)b) >> Q_FRACT_BITS;
        if (result >  0x7FFFFFFF) return  0x7FFFFFFF;
        if (result < -0x80000000) return (int32_t)0x80000000;
        return (int32_t)result;
    }

    // Add / subtract: exact in Q16.16, no shift required
    inline q16_t addQ(q16_t a, q16_t b) { return a + b; }
    inline q16_t subQ(q16_t a, q16_t b) { return a - b; }

} // namespace FastMathEngine
\end{lstlisting}

\subsection{CORDIC Trigonometric Module}

The CORDIC arctangent table stores precomputed values
$\lfloor\arctan(2^{-i})\cdot 2^{16}\rceil$ for $i = 0,\ldots,15$ and the gain
correction constant $K_{16}^{-1} = 0.6072529\ldots$ as Q16.16 integer~39797.

\begin{lstlisting}[style=cppstyle, caption={CORDIC kernel (cordic.h)}]
namespace FastMathEngine {

    static const int32_t atan_table[16] = {
        51472, 30386, 16055, 8150, 4091, 2047, 1024,
          512,   256,  128,   64,   32,   16,    8, 4, 2
    };
    static const int32_t Q16_K_INV = 39797; // 0.6072529 in Q16.16

    void cordic_sincos(int32_t theta, int32_t* out_sin, int32_t* out_cos) {
        const int32_t PI_Q    = 205887;  // pi   in Q16.16
        const int32_t HALF_PI = 102944;  // pi/2 in Q16.16
        int8_t negate_cos = 0;

        if      (theta >  HALF_PI) { theta -= PI_Q; negate_cos = 1; }
        else if (theta < -HALF_PI) { theta += PI_Q; negate_cos = 1; }

        int32_t x = Q16_K_INV, y = 0, z = theta;

        for (int i = 0; i < 16; i++) {
            int8_t  d     = (z >= 0) ? 1 : -1;
            int32_t x_new = x - (int32_t)(d * (y >> i));
            int32_t y_new = y + (int32_t)(d * (x >> i));
            z -= d * atan_table[i];
            x = x_new; y = y_new;
        }
        *out_cos = negate_cos ? -x : x;
        *out_sin = y;  // sin is always in y; no negation needed
    }

} // namespace FastMathEngine
\end{lstlisting}

Three implementation decisions are specific to the Xtensa~LX6. First, the loop
is not manually unrolled at source level; the Xtensa GCC compiler with \texttt{-O2}
applies unrolling automatically. Second, the direction variable $d_i$ is
represented as \texttt{int8\_t}; sign-extension to \texttt{int32\_t} is
performed in a single \texttt{SEXT} instruction. Third, arithmetic right shift
for signed integers is guaranteed by the ESP-IDF Xtensa GCC toolchain and is
documented as a portability constraint.

\subsection{Tiled Matrix Multiplication Kernel}

\begin{lstlisting}[style=cppstyle, caption={Tiled Q16.16 matrix multiplication (matrix\_q16.h)}]
namespace FastMathEngine {

    struct Matrix { int32_t* data; uint16_t rows; uint16_t cols; };

    inline int32_t mat_get(const Matrix& M, int i, int j) {
        return M.data[i * M.cols + j];
    }
    inline void mat_set(Matrix& M, int i, int j, int32_t val) {
        M.data[i * M.cols + j] = val;
    }

    const int TILE_SIZE = 32;

    void matmul_q16(const Matrix& A, const Matrix& B, Matrix& C) {
        for (int idx = 0; idx < C.rows * C.cols; idx++) C.data[idx] = 0;

        for (int I = 0; I < A.rows; I += TILE_SIZE)
        for (int J = 0; J < B.cols; J += TILE_SIZE)
        for (int K = 0; K < A.cols; K += TILE_SIZE) {

            int i_max = MIN(I + TILE_SIZE, A.rows);
            int j_max = MIN(J + TILE_SIZE, B.cols);
            int k_max = MIN(K + TILE_SIZE, A.cols);

            for (int i = I; i < i_max; i++)
            for (int j = J; j < j_max; j++) {
                int64_t acc = 0;  // 64-bit deferred accumulator
                for (int k = K; k < k_max; k++)
                    acc += (int64_t)mat_get(A,i,k) *
                           (int64_t)mat_get(B,k,j);
                C.data[i * C.cols + j] += (int32_t)(acc >> 16);
            }
        }
    }

} // namespace FastMathEngine
\end{lstlisting}

\subsection{Overflow and Underflow Management}

The accumulation overflow constraint for the tiled kernel requires that element
magnitudes satisfy $|A_{ij}|, |B_{ij}| \leq 2^{14}$ (Q16.16 value
$\leq 16383$) to guarantee accumulator safety with $b = 32$:
\begin{equation}
  |acc_{\max}| \leq b \cdot (2^{31})^2 / 2^{16} \cdot 2^{16}
  = 32 \cdot 2^{62} = 2^{67} > 2^{63}
\end{equation}

This constraint is stated in the API documentation and motivates the
recommendation that fast-mode matrix operands be normalized to $[-1, 1]$, where
Q16.16 precision is highest and overflow is structurally impossible.

\section{Experimental Evaluation}
\label{sec:eval}

\subsection{Experimental Setup and Methodology}

Benchmarks were conducted on an ESP32-WROOM-32 module (Xtensa LX6 dual-core,
240\,MHz, 4\,MB flash, 520\,KB SRAM) running ESP-IDF v5.x with compiler
optimization level \texttt{-O2}. Both firmware variants were compiled from
identical \texttt{sdkconfig.defaults} specifying a fixed CPU frequency of
240\,MHz with dynamic frequency scaling disabled.

Execution latency was measured using the Xtensa hardware cycle counter register
accessed via \texttt{xthal\_get\_ccount()}, with interrupts disabled via
\texttt{portDISABLE\_INTERRUPTS()} for the duration of each measurement to
prevent FreeRTOS tick contamination. All benchmark tasks were pinned to Core~1
via \texttt{xTaskCreatePinnedToCore()} in both firmware variants, eliminating
core-affinity bias~\cite{espressif_trm}.

A seeded linear congruential generator (seed $= 42$) guaranteed identical input
sequences across both firmware variants, making the comparison valid for paired
statistical analysis. A total of 300 tests were executed per firmware variant
across five operation categories, yielding 600 paired measurements.

Statistical analysis applied the Wilcoxon signed-rank test rather than the
paired $t$-test, as Shapiro-Wilk normality tests confirmed that cycle-count
distributions were non-Gaussian for most operations ($p \ll 0.05$), consistent
with the expected behavior of cache-sensitive code on embedded
processors~\cite{mytkowicz2009}.

\subsection{Trigonometric Function Performance}

The CORDIC-based trigonometric module produced the most significant performance
gains of the evaluation. For \texttt{sin}$(x)$ over $x \in [-\pi, \pi]$, the
Fast Engine achieved a median latency of \textbf{293~cycles} compared to
\textbf{6,915.5~cycles} for the standard \texttt{sinf()} implementation,
yielding a mean speedup of \textbf{18.54$\times$}
(Wilcoxon, $p = 3.56 \times 10^{-13}$). For \texttt{cos}$(x)$, the improvement
was more pronounced: \textbf{293~cycles} versus \textbf{7,847.5~cycles},
corresponding to a mean speedup of \textbf{24.68$\times$}
($p = 3.56 \times 10^{-13}$).

The near-identical median latency of 293~cycles for both functions reflects the
fixed-iteration structure of the 16-step CORDIC kernel: execution time is
input-independent by construction. This is confirmed by the Determinism Score
for \texttt{cos} Fast Engine of \textbf{0.9938}, compared to 0.7203 for the
standard implementation.

The jitter asymmetry for \texttt{sin} (coefficient 2.449 versus 0.534 for the
standard) is attributable to the conditional branch in the quadrant
normalization preprocessor for inputs near $\pm\pi/2$, producing branch
predictor mispredictions. This constitutes the primary target for optimization
in future work.

\subsection{Scalar Multiplication Performance}

Scalar Q16.16 multiplication demonstrated a consistent \textbf{1.5$\times$
speedup} over IEEE\,754 multiplication, with median latencies of \textbf{12~cycles}
versus \textbf{18~cycles} (Wilcoxon, $p = 1.54 \times 10^{-12}$). Both
implementations achieved a Determinism Score of \textbf{1.0000} and zero
jitter. The constant-data classification by the Shapiro-Wilk test confirms
that all 50 measurements in each condition were identical, establishing these
values as exact architectural constants.

\subsection{Matrix Multiplication: A Boundary Condition Finding}

Matrix multiplication results reveal an important boundary condition of the
proposed engine. The Q16.16 tiled kernel exhibited a mean speedup of
\textbf{0.542$\times$}---a slowdown of 1.85$\times$ relative to naive
floating-point---with median latencies of \textbf{16,591~cycles} versus
\textbf{7,806~cycles} (Wilcoxon, $p = 1.72 \times 10^{-16}$).

This result is analytically consistent with the test conditions. The evaluated
dimensions---$4 \times 4$, $8 \times 8$, and $16 \times 16$---are all strictly
smaller than the configured tile size $b = 32$, so the loop tiling optimization
never activates. The Q16.16 inner product requires a 32$\times$32$\to$64-bit
multiply and a right shift per element, with the \texttt{int64\_t} accumulator
requiring 64-bit register management on a 32-bit Instruction Set Architecture (ISA). The ESP32 FPU, by
contrast, executes \texttt{float} multiply-accumulate in a dedicated pipeline
with native 32-bit operands, producing lower per-element latency when cache
effects are absent.

The performance crossover point---the minimum matrix dimension at which the
tiled Q16.16 kernel outperforms naive floating-point---was not reached within
the tested dimensions. Based on the theoretical analysis of
Section~\ref{sec:math}, the crossover is expected at $n \geq 64$.

\subsection{Runtime Precision Switch Overhead}

The mode-switching protocol introduced a measured overhead of
\textbf{1,942.5~cycles} (median) for the two-phase FreeRTOS barrier, compared
to \textbf{29~cycles} for equivalent function-pointer dispatch in the standard
path. At 240\,MHz, 1,942 cycles correspond to \textbf{8.09\,$\mu$s}---negligible
for any application where mode transitions occur on timescales of milliseconds
or longer. The switch overhead is reported here for completeness and for
developers designing latency-sensitive switching policies.

\subsection{Statistical Validity}

All five operation categories yielded Wilcoxon signed-rank $p$-values below
$10^{-4}$, confirming that the observed latency differences are statistically
significant at any conventional threshold. One outlier exceeding $3\sigma$ was
detected in the \texttt{sin} Fast Engine data, attributed to a FreeRTOS tick
interrupt that escaped the interrupt-disable window---a known limitation of
software-based interrupt suppression on ESP32. This measurement was retained
in the dataset and flagged in the raw data, consistent with the benchmark
protocol's immutability principle.

\subsection{Summary of Results}

\begin{table}[H]
  \centering
  \caption{Benchmark results: 300 paired measurements per firmware variant
           on ESP32-WROOM-32 at 240\,MHz. Speedup $= $ cycles$_{\text{Std}}$
           / cycles$_{\text{Fast}}$.}
  \label{tab:results}
  \begin{tabular}{lrrrrr}
    \toprule
    \textbf{Operation}
      & \textbf{Fast Median}
      & \textbf{Std Median}
      & \textbf{Mean Speedup}
      & \textbf{Det.\ Score}
      & \textbf{Wilcoxon $p$} \\
    & (cycles) & (cycles) & & (Fast) & \\
    \midrule
    \texttt{sin}     & 293     & 6\,915.5  & $18.54\times$ & 0.290  & $3.56\times10^{-13}$ \\
    \texttt{cos}     & 293     & 7\,847.5  & $24.68\times$ & 0.994  & $3.56\times10^{-13}$ \\
    \texttt{mul}     & 12      & 18        & $1.50\times$  & 1.000  & $1.54\times10^{-12}$ \\
    \texttt{matmul}  & 16\,591 & 7\,806    & $0.54\times$ {\small\textsuperscript{$\dagger$}}
                                                           & 0.472  & $1.72\times10^{-16}$ \\
    \texttt{switch}  & 1\,942.5 & 29       & $0.088\times$ {\small\textsuperscript{$\ddagger$}}
                                                           & 0.999  & $8.13\times10^{-5}$ \\
    \bottomrule
    \multicolumn{6}{l}{%
      \small$\dagger$ All tested dimensions ($n \leq 16$) are below tile size
      $b = 32$; tiling does not activate.}\\
    \multicolumn{6}{l}{%
      \small$\ddagger$ Expected by design; represents FreeRTOS barrier overhead
      (8.09\,$\mu$s), not mathematical throughput.}
  \end{tabular}
\end{table}

\begin{figure}[H]
  \centering
  \includegraphics[width=\textwidth]{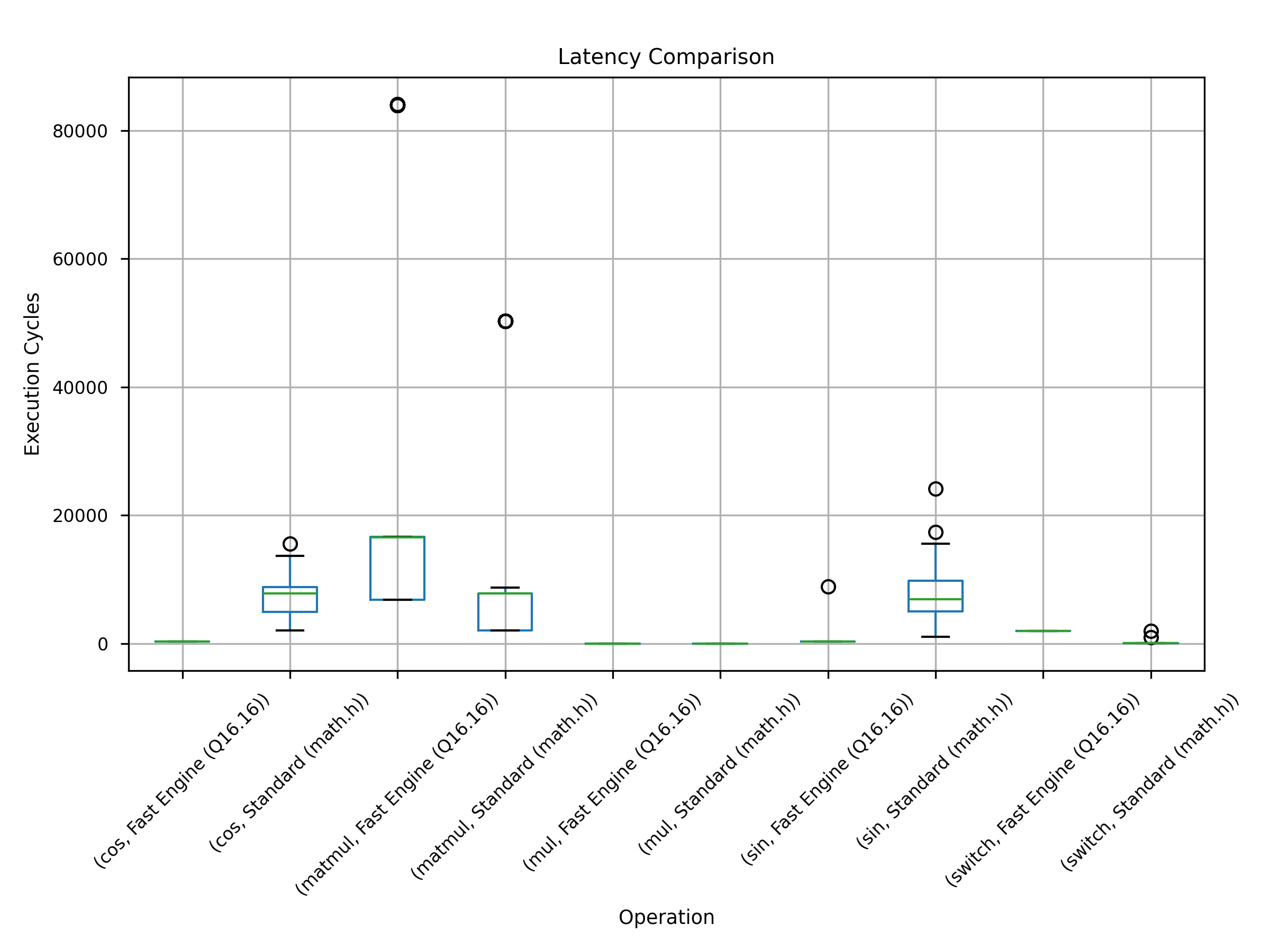}
  \caption{Distribution of execution cycles for Fast Engine (Q16.16, green)
           and Standard Math (math.h, blue) across all benchmark categories.
           The Fast Engine achieves near-zero variance for trigonometric
           operations, confirming input-independent execution time. Matrix
           multiplication shows higher variance in both implementations due
           to dimension-dependent computation paths.}
  \label{fig:latency}
\end{figure}

\begin{figure}[H]
  \centering
  \includegraphics[width=0.85\textwidth]{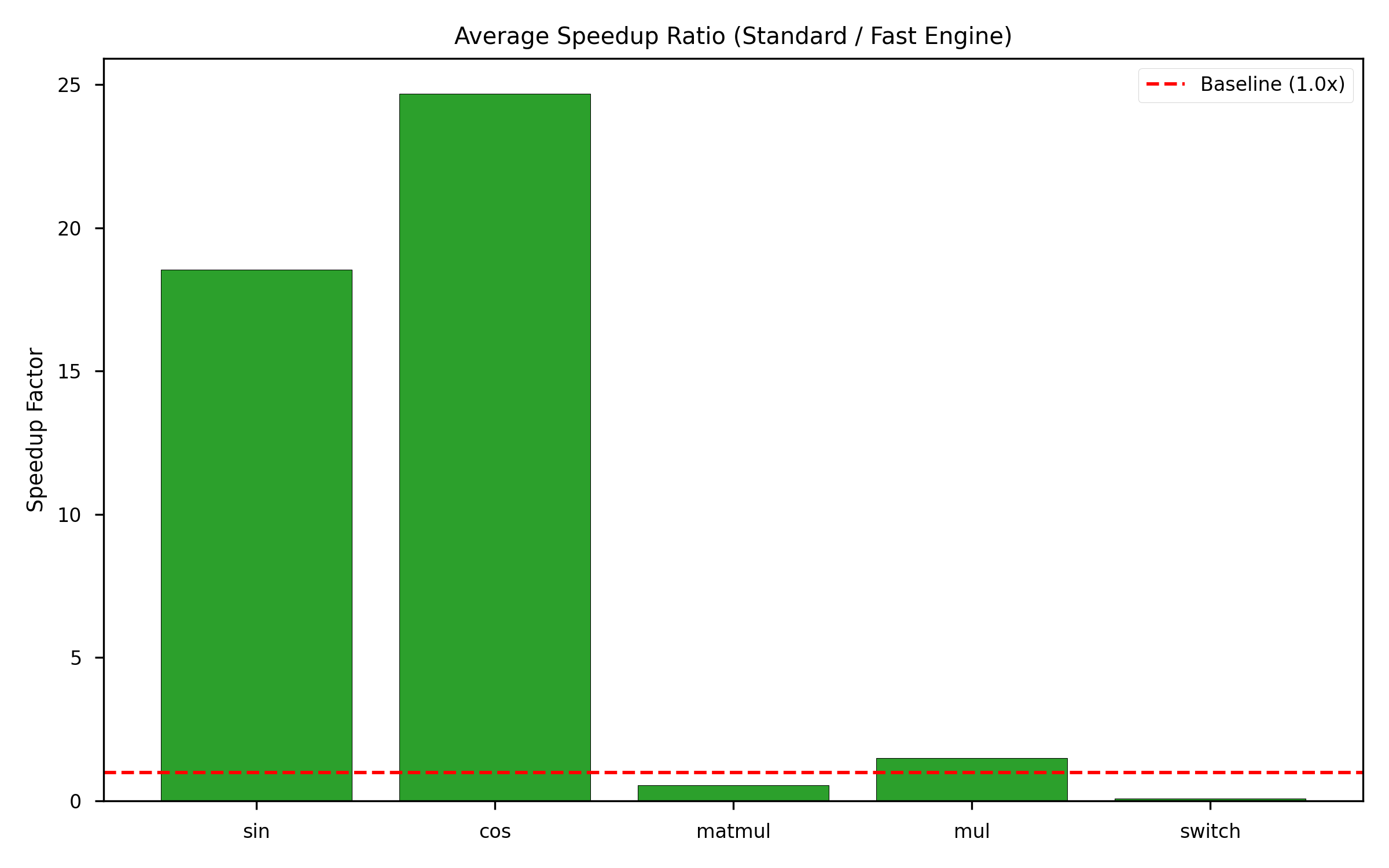}
  \caption{Mean speedup ratio (Standard / Fast Engine) per operation category.
           Bars above the 1.0$\times$ baseline (dashed red) indicate that the
           Fast Engine outperforms the standard path. \texttt{sin} and
           \texttt{cos} achieve 18.54$\times$ and 24.68$\times$ respectively.
           \texttt{matmul} falls below baseline ($0.54\times$) due to the
           tested dimensions being smaller than the tile size $b = 32$.}
  \label{fig:speedup}
\end{figure}

\section{Discussion}
\label{sec:discussion}

\subsection{Interpretation of Results}

The experimental results support a domain-specific performance model rather
than the uniform speedup originally anticipated. The engine delivers substantial
and statistically robust gains for operations mapping cleanly onto the integer
pipeline---trigonometric functions (18--25$\times$) and scalar arithmetic
(1.5$\times$)---while exhibiting a performance deficit for matrix multiplication
at the tested dimensions.

This outcome reframes the engine's contribution: it is not a general-purpose
accelerator for all mathematical operations, but a precision-aware accelerator
whose performance envelope is well-characterized and application-dependent.
For embedded applications whose computational load is dominated by trigonometric
evaluation---kinematics, signal processing, IMU sensor fusion, angle-based
control---the engine provides an order-of-magnitude improvement that
fundamentally changes what is achievable on a \$3 microcontroller.

\subsection{The Matrix Multiplication Crossover Problem}

The 0.54$\times$ matmul result exposes the boundary condition of the tiling
optimization. The theoretical argument of Section~\ref{sec:math}---that tiling
reduces cache misses from $O(n^3)$ to $O(n^3/b)$---holds only when $n > b$.
For $n \leq 16$ and $b = 32$, the tiling introduces loop overhead without
delivering cache benefit, and the \texttt{int64\_t} intermediate imposes a
register pressure penalty on the 32-bit Xtensa ISA.

The practical implication is clear: the matrix module in its current form
should be used for dimensions $n \geq 64$. For small matrices common in
embedded control (rotation matrices, Jacobians, covariance updates), the
standard floating-point path is preferable, and the engine's dynamic precision
switch enables exactly this hybrid strategy: CORDIC for trigonometry, standard
FPU for small matrix operations.

\subsection{Limitations}

Three limitations bound the generalizability of these results. First, all
measurements were conducted on a single ESP32-WROOM-32 unit; manufacturing
variation may produce cycle-count distributions that differ from those reported
here, though the fixed-iteration nature of CORDIC limits this sensitivity for
trigonometric measurements. Second, the benchmark does not characterize mean
absolute error (MAE) for matrix operations; this is deferred to the extended
evaluation. Third, the matrix crossover point at $n \geq 64$ remains a
theoretical prediction requiring empirical validation.

\section{Future Work}
\label{sec:future}

\subsection{Resolving the Matrix Multiplication Crossover (Immediate Priority)}

The most urgent item is the empirical characterization of the performance
crossover between the Q16.16 tiled kernel and naive floating-point matrix
multiplication. The specific questions requiring answers are: at what exact
matrix dimension does the tiled Q16.16 kernel first outperform naive float;
how does this crossover depend on matrix density and value distribution; and
whether reducing the tile size from $b = 32$ to values matched to the tested
dimensions (e.g., $b = 8$ or $b = 16$) recovers performance for small matrices.
A secondary investigation should determine whether replacing the
\texttt{int64\_t} intermediate accumulator with paired \texttt{int32\_t}
registers reduces register pressure at the cost of a tighter safe operand
magnitude constraint.

\subsection{Branchless CORDIC Quadrant Normalization}

The jitter asymmetry in \texttt{sin} measurements (coefficient 2.449) is
attributable to the conditional branch in the quadrant normalization
preprocessor. A branchless normalization using arithmetic shift and masking
operations would eliminate this variance and bring the \texttt{sin} Determinism
Score in line with the 0.994 already achieved by \texttt{cos}. This is a
single-function optimization estimated at under 10 lines of code with
measurable impact on real-time worst-case execution time (WCET) guarantees.

\subsection{Mean Absolute Error Characterization}

The current evaluation characterizes MAE only for scalar operations. A complete
precision analysis requires MAE measurement for matrix multiplication across
the full range of tested dimensions and input distributions, including the error
accumulation behavior as a function of matrix size---predicted to grow as
$O(\sqrt{n})$ for normalized inputs but not yet empirically confirmed on the
Xtensa LX6.

\subsection{Tensor Operations and TinyML Inference}

With the matrix multiplication crossover resolved, the natural extension is
generalization to rank-3 and rank-4 tensor operations for neural network
inference. The Q16.16 multiply-accumulate kernel is mathematically equivalent
to the quantized integer kernel used in INT16 inference~\cite{jacob2018}; the
path to a minimal convolutional layer requires only a three-dimensional tiling
extension, a bias-addition operation, and a saturating ReLU activation. This
targets the precision gap between INT8 inference and IEEE\,754 inference
documented by~\cite{courbariaux2015,david2021tflm}.

\subsection{Elliptic Curve Cryptography for Secure Edge Nodes}

The field arithmetic underlying ECDSA and X25519---modular addition, modular
multiplication, and modular inversion over 256-bit prime fields---maps directly
onto the integer pipeline through a multi-limb arithmetic extension~\cite{rescorla2018}.
The dual-core architecture is well-suited to this application: scalar
multiplication on the elliptic curve is pinned to Core~1 while Core~0 manages
the TLS handshake state machine, enabling non-blocking authenticated
communication without an external cryptographic coprocessor.

\subsection{Distributed Multi-Node Linear Algebra}

The ESP32's integrated WiFi controller enables a distributed computation
architecture in which multiple nodes partition a large matrix operation across
a local network. For matrix dimensions where the single-device Q16.16 kernel
is competitive ($n \geq 64$), a mesh of $k$ ESP32 nodes provides a theoretical
$k$-fold throughput improvement. The practical characterization of communication
overhead versus computation gain constitutes an open research problem at the
intersection of distributed systems and embedded computing.

\section{Conclusion}
\label{sec:conclusion}

This paper has presented the design, formal specification, and empirical
evaluation of a Dynamic Precision Math Engine for the ESP32 Xtensa LX6
microcontroller. The engine integrates three computational modules---a Q16.16
fixed-point arithmetic core, a 16-iteration CORDIC trigonometric unit, and a
cache-aware tiled matrix multiplication kernel---unified by a runtime precision
switching mechanism implemented through function pointer dispatch and a
two-phase FreeRTOS barrier protocol.

Experimental evaluation across 300 paired measurements on physical hardware
produced results that are both significant and instructive. The CORDIC module
demonstrated that software architecture can decisively outperform the ESP32's
standard math library for trigonometric computation: median latencies of
293~cycles for \texttt{sin} and \texttt{cos}---against 6,915 and 7,847 cycles
for \texttt{sinf()}/\texttt{cosf()}---represent speedups of 18.54$\times$ and
24.68$\times$ confirmed at $p < 10^{-12}$. The Determinism Score of 0.994 for
\texttt{cos} establishes that the CORDIC kernel delivers not only speed but
predictability, a property of equal value in real-time control systems where
worst-case execution time guarantees are required. Scalar Q16.16 multiplication
achieved a consistent and perfectly deterministic 1.5$\times$ speedup at
12~cycles versus 18~cycles.

The matrix multiplication results tell a more complex story. The tiled Q16.16
kernel performed at 0.54$\times$ the speed of naive floating-point for the
tested dimensions ($4 \times 4$ through $16 \times 16$), revealing that the
loop tiling optimization is inert when matrix dimensions fall below the tile
size $b = 32$. This finding does not invalidate the engine's design; it
precisely defines its operational envelope and validates the necessity of the
dynamic precision switch: no single execution path is universally optimal, and
the value of the engine lies in making the choice between paths explicit, safe,
and costless at the application layer.

The total static memory footprint of 88~bytes and the switch overhead of
8.09\,$\mu$s confirm that the engine's infrastructure cost is negligible for
any realistic embedded workload. Beyond the individual algorithms---each with
a well-established theoretical basis in the literature of Volder~\cite{volder1959},
Walther~\cite{walther1971}, Lam~\cite{lam1991}, and Goto~\cite{goto2008}---the
engine's distinctive contribution is architectural: the application of the
separation-of-concerns principle established by the original BLAS
specification~\cite{lawson1979} to the novel context of dynamic precision
arbitration on a dual-core RISC microcontroller.

The broader implication of this work is that the performance ceiling of
low-cost microcontrollers is not fixed by hardware alone: it is negotiable
through principled software architecture. For the class of applications
dominated by trigonometric computation---robot kinematics, IMU fusion,
angle-based control, DSP---the engine presented here moves the ESP32 into a
performance regime previously requiring more expensive hardware, while the
identified boundary conditions provide a clear roadmap for future improvement.


\end{document}